\documentclass{aa}

\input psfig

\begin{document}
\title{The Parkes quarter-Jansky flat-spectrum sample}
\subtitle{2. New Optical Spectra and Redshift Measurements}
\author{I. M. Hook\inst{1}  \and 
        P. A. Shaver\inst{2} \and  
        C. A. Jackson\inst{3} \and 
        J. V. Wall\inst{1} \and 
        K. I. Kellermann\inst{4}}

\institute{
 Department of Physics, University of Oxford,
 Nuclear \& Astrophysics Laboratory,
 Keble Road,
 Oxford,  OX1 3RH, UK
\and European Southern Observatory, Karl Schwarzschild Stra\ss{e} 2, 
     D-85748 Garching b. M\"{u}nchen, Germany
\and Research School of Astronomy \& Astrophysics, 
       The Australian National University, Mount Stromlo Observatory,
       Canberra, Australia
\and National Radio Astronomy
 Observatory, 520 Edgemont Road, Charlottesville, VA 22903-2475, USA\thanks{The National Radio Astronomy Observatory is operated by Associated
Universities Inc. under a cooperative agreement with the National
Science Foundation. }}

\offprints{I. M. Hook \email{imh@astro.ox.ac.uk}}

\date{version \today}

\abstract{We present optical spectra and redshift measurements for 178
flat-spectrum objects from the Parkes quarter-Jansky flat-spectrum
sample\thanks{Table A1 is also available in electronic form at the CDS
via anonymous ftp to {\tt cdsarc.u-strasbg.fr} (130.79.128.5) or via
{\tt http://cdsweb.u-strasbg.fr/cgi-bin/qcat?J/A+A/}}. These spectra
were obtained in order to compile a complete sample of quasars for use
in a study of quasar evolution.  We present a composite optical
spectrum made from the subset of 109 quasars that have flux densities
in the range $\rm 0.25Jy < S_{2.7GHz}< 0.5Jy$, and make a comparison
with a composite for radio-quiet QSOs from the Large Bright Quasar
Survey.  Our large sample of radio-loud quasars allows us to
strengthen previous reports that the Ly$\alpha$ and CIV emission lines
have larger equivalent width in radio-loud quasars than radio-quiet
QSOs to greater than the 3$\sigma$ level.  However we see no
significant difference in the equivalent widths of CIII] or MgII.  We
also show that the flux decrements across the Lyman-$\alpha$ line
($D_A$) measured from these spectra show the same trend with redshift
as for optically selected QSOs.
}
\maketitle

\section{Introduction}

This is the second paper in a series of three describing the results
of a program to search for high-redshift quasars and study the
evolution of the flat-spectrum radio-loud quasar population.  The
first paper (Paper~1, Jackson et al. 2002) set out the sample,
discussing its selection and the identification and re-confirmation
programme to determine the optical counterparts to the radio sources.
This paper, Paper~2, presents new spectroscopic observations and
redshift determinations.  Paper~3 (Wall et al. 2002) considers the
quasar space distribution, the luminosity function and its epoch
dependence.  Because the sample is radio selected, the completeness of
the sample is unaffected by obscuration from dust. An earlier version
of this sample was used by Shaver et al. (1996b) to show that the
space density of quasars declines at redshifts greater than~3.

In addition to the primary goal of this work (the study of quasar
evolution), the sample can be used for several other statistical
studies. It has provided the basis for a study of damped
Lyman-$\alpha$ absorption systems along the lines of sight to the
quasars (Ellison et al. 2001, 2002). In this paper we also consider
some properties of the quasar spectra themselves, in particular the
equivalent width distributions of the emission lines, and the flux
decrement across the Lyman-$\alpha$ line at high redshift.

\section{Definition of the spectroscopic sample}

We have defined a sample of flat-spectrum radio sources from the
PKSCAT90 catalogue (Wright \& Otrupeck 1990) as described in Jackson
et al. (2002). The sample of 878 sources was selected based on the
following criteria: $\alpha_{2.7 \rm \thinspace GHz}^{5.0 \rm
\thinspace GHz} > -0.4$, where $S \propto \nu ^{\alpha}$, $-80^\circ <
\delta < +2.5^\circ$, $|b|\ge 10^\circ$. The optical identification
procedure, including new radio positions that were obtained as part of
this study, is described in Jackson et al. (2002) and a full set of
finding charts is given there. The majority of identifications were
made using COSMOS scans of UKST $B_J$ plates, and CCD identifications
were obtained for those sources not identified on the plates.  We
refer to this sample as the Parkes quarter-Jansky flat-spectrum
sample.

Since the primary aim of this project was to study the evolution of
the quasar population using a complete sample, we then selected for
spectroscopic follow up all those stellar sources (as determined from
COSMOS or CCD identifications) with $\rm S_{2.7GHz}\ge 0.25Jy$ in the
zone north of $-45^{\circ}$ that did not have previous redshift
measurements in the literature. In addition, a few sources south of declination $-45^{\circ}$
were also observed and their spectra are included in this paper,
although the sample was not intended to be complete in that region.

\section{Spectroscopic Observations}
The spectroscopic observations were carried out at the ESO 3.6m at La
Silla, Chile during observing runs in November 1993, May 1994, October
1994, April 1995, October 1995, May 1997 and October 1997.
Observations were made of a total of 178 sources, the majority during
the last two of these runs. Low resolution spectra were obtained using
the EFOSC spectrograph with a 1.5'' slit width. Most sources were
observed at a sky position angle of 270$^\circ$, i.e. not necessarily
at the parallactic angle. In most cases the B300 grism (wavelength
range 3750-6950\AA, dispersion 6.35\AA\ per pixel) was used, but a few
sources were observed with the R300 grism (wavelength range
6000-9910\AA, dispersion 7.7\AA\ per pixel) either because the object
was clearly red based on CCD identifications, or because the blue
spectrum was ambiguous.

\section{Results}

As a result of the spectroscopic observations described above, the
Parkes quarter-Jansky flat-spectrum sample has highly complete
redshift information (see Jackson et al. 2002).  In
summary, there are 677 objects classified as quasars or BLs in the
original sample of 878. A total of 174 of the objects classified as
'Q' (and none of the BLs) do not have spectroscopic confirmation. Note
that in the analysis of the quasar space density in Paper~3, only
the near-complete sub-sample with $\delta > -40\deg$ is considered.

Of the 178 objects that we observed, 126 show strong, broad emission
lines and were therefore classified as quasars. 37 objects showed
featureless spectra or very weak lines on a strong continuum, and we
classified these as BL Lac objects. The remaining 15 objects were
classified as galaxies based on the presence of galaxy absorption
lines, narrow emission lines or a strong 4000\AA\ break.

Mean redshifts were calculated from the emission lines (or absorption
lines in a few cases). The rest frame wavelengths used for the quasar
emission lines were those given in Tytler \& Fan (1992), which have
been corrected for small ($\sim 0.1\%$) systematic shifts between UV
emission lines and the true systemic velocity. Since the Ly-$\alpha$
line is usually affected by Ly-$\alpha$ forest absorption it was not
used to calculate the redshift unless there were no other clean lines
in the spectrum.

\begin{figure*}
\centering
\mbox{\psfig{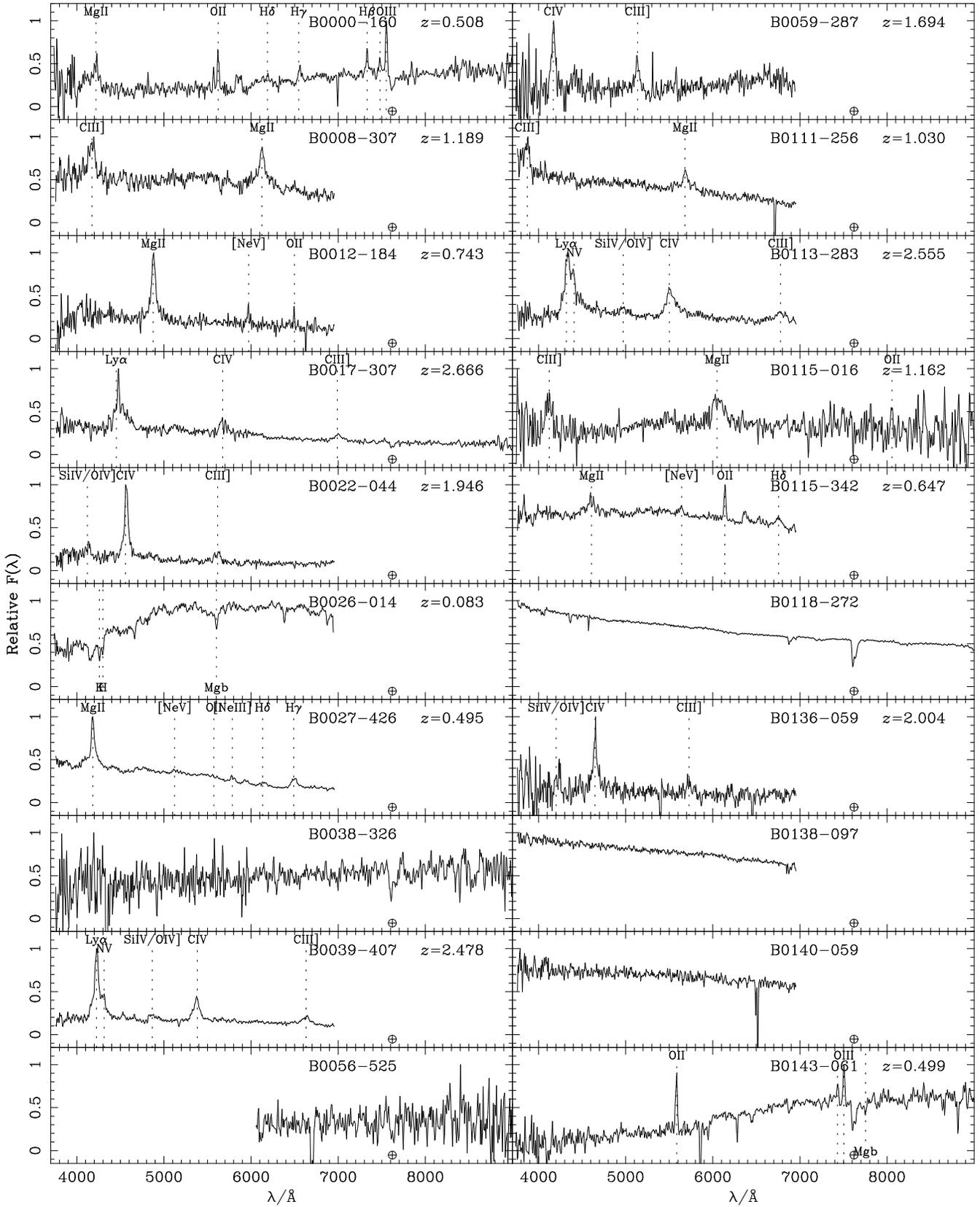}}
\caption[]{
New optical spectra for the Parkes quarter-Jansky flat-spectrum
sample.  The spectra have been scaled to have a maximum flux value of
1.0.  The dotted vertical lines show the expected positions of
spectral features at the measured mean redshift (not the measured
wavelengths of the individual features). An asterisk after the
redshift indicates an uncertain redshift.  The earth symbol at $\sim
7600$\AA\ indicates the position of the atmospheric absorption feature
(the A-band).}
\label{specfig}
\end{figure*}

\begin{figure}
\centering
\mbox{\psfig{figure=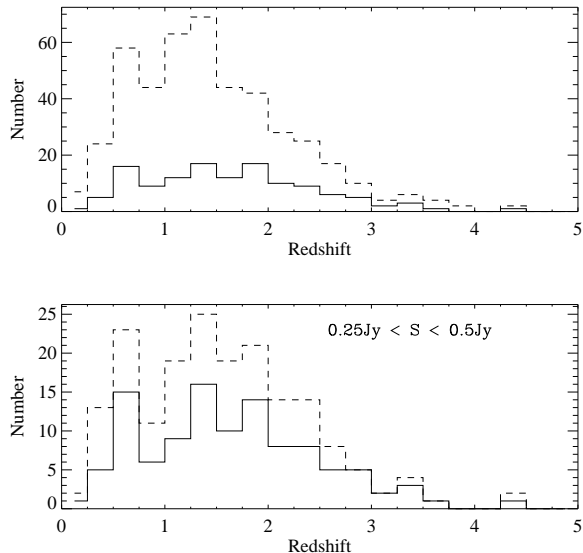,height=2.9in,bbllx=36pt,bblly=12pt,bburx=533pt,bbury=489pt}}
\caption[]{Upper panel: Redshift histogram for all 449 quasars in the Parkes quarter-Jansky flat-spectrum sample (dashed line) and
for the 126 new quasar spectra presented in this paper (solid
line). Lower panel: As above but for sources with $\rm 0.25Jy \le S <
0.5Jy$ (183 and 109 sources respectively).}
\label{pks_zhist}
\end{figure}

\begin{figure}
\centering
\mbox{\psfig{figure=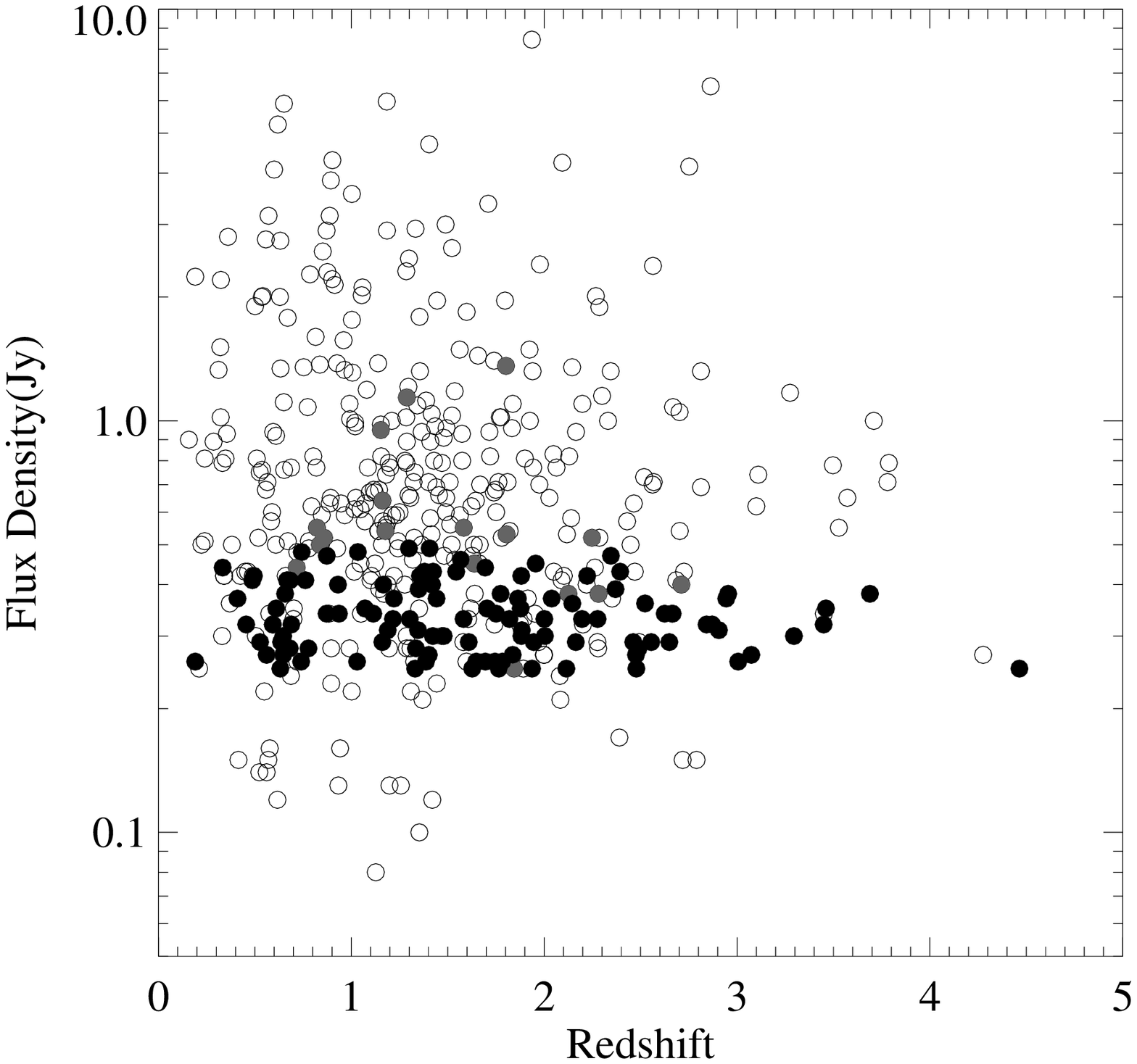,height=2.7in,bbllx=29pt,bblly=14pt,bburx=528pt,bbury=485pt}}
\caption[]{Flux density vs  Redshift for all quasars in the Parkes quarter-Jansky flat-spectrum sample (open circles). 
Quasars with new redshifts presented in this paper are plotted as
filled circles: black circles for those used to make the composite,
grey for the others.}
\label{sz}
\end{figure}

The new spectra are plotted in Figure~\ref{specfig} and the redshifts
are given in Table~\ref{pksz}.  The redshift distribution for the new
quasar spectra is shown in Figure~\ref{pks_zhist}. The flux density
distribution for these objects is also plotted as a
function of redshift in Figure~\ref{sz}.

The vast majority (109 from 126) of the quasars for which we have
obtained spectra have flux densities in the range $\rm
0.25Jy<S_{2.7}<0.5Jy$ (almost all sources stronger than this had
redshifts from the literature). A K-S test shows that the distribution
of these 109 new quasar redshifts is indistinguishable from that that
of all quasars in the Parkes quarter-Jansky flat-spectrum sample with
$\rm 0.25Jy<S_{2.7}<0.5Jy$.  In the remainder of this paper we take
this sample of 109 new quasar spectra as representative of the Parkes
quarter-Jansky flat-spectrum quasar sample in this flux range, and we
term this sample the PKS 0.25-0.5Jy sample.

\subsection{Notes on Individual Objects} 

{\bf B0039$-$407} Iovino, Clowes \& Shaver (1996) estimated the
redshift as $z=2.37$ from an objective prism measurement. We now
measure $z=2.478$.

{\bf B0138$-$097} Stocke \& Rector (1997) give $z=0.733$ for this BL
source. Despite high signal-to-noise in the range
$4000-7000\AA$, our spectrum does not confirm their MgII emission-line
redshift of $z=0.733$. We classify this source as a BL with
undetermined redshift.

{\bf B0357$-$264} Drinkwater et al. (1997) give ``$z$=1.47?'' for this
source.  Their spectrum does not show any strong features and
it is not clear from their paper which features this redshift is based
on. Our spectrum covers a similar wavelength range to that of the
Drinkwater et al. spectrum but does not show sufficient features to
measure a redshift. We therefore classify this source as a BL. 

{\bf B0646$-$306} Veron-Cetty \& Veron (1993) give $z$=0.455 for this
object whereas we find $z=1.153$. It is likely that the previous
redshift determination was the result of misidentifying the CIII] line
at $\rm \sim 4100\AA$ with MgII.

{\bf B1136$-$156} The emission lines of CIV and Ly$\alpha$ are
affected by absorption.

{\bf B1251$-$407} The previously published redshift of 4.458 (Shaver
at al 1996a) was derived using different assumptions for the
rest-frame wavelengths of the emission-lines. Using values corrected
for systematic shifts (Tytler \& Fan 1992) we derive a redshift of
4.464.

{\bf B2303$-$656} PKSCAT90 lists this object at a redshift of
$z=0.47$.  Our spectrum, which starts at $\rm 6000\AA$, does not show
any clear features to confirm this.

{\bf B2224$+$006} The emission lines of CIV and Ly$\alpha$ are
affected by absorption.

\section{A Composite radio-selected QSO spectrum}

\subsection{Method for producing the composite spectrum}

We have derived a composite spectrum from 109 of the quasar spectra
shown in Figure~\ref{specfig} (i.e. those that lie in the PKS
0.25-0.5Jy sample) with the aim of comparing emission line strengths
for radio and optically selected QSOs.

The B300 spectra were clipped to cover $3750 < \lambda < 6830$\AA\ and
those spectra taken with the R300 grism were clipped to $\lambda <
9300$\AA.  The region from $7595-7645$\AA\ in the R300
spectra, which is strongly affected by the sky absorption feature
(A-band), was cut out of the spectra.

\begin{figure*}
\centering
\mbox{\psfig{figure=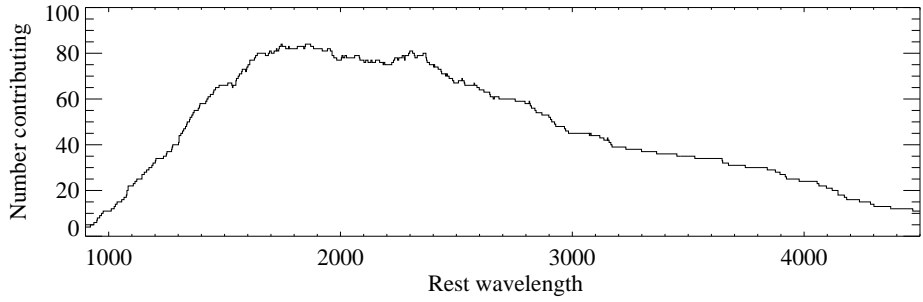,height=1.75in,bbllx=0pt,bblly=550pt,bburx=510pt,bbury=722pt}}
\caption[]{The
number of spectra contributing to the PKS 0.25-0.5Jy composite as a function of
wavelength.}
\label{makecomp}
\end{figure*}

The spectra were then sorted by redshift, shifted to the rest frame
and interpolated onto a common rest-frame wavelength grid. The
composite was built up iteratively by including one more spectrum at
each iteration. A scale factor was calculated for each spectrum by
normalising it to the current composite (since the spectra were sorted
in redshift, there is a large region of overlap in the rest frame
between each new spectrum and the most recent iteration of the
composite). Each iteration of the composite was formed by adding all
the contributing points at that wavelength, scaled by the appropriate
scale factor for that spectrum, and divided by sum of scale factors.
Thus each spectrum has an equal weight in the composite.
Figure~\ref{makecomp} shows the number of spectra contributing to each
rest wavelength bin.

\begin{figure*}
\centering
\mbox{\psfig{figure=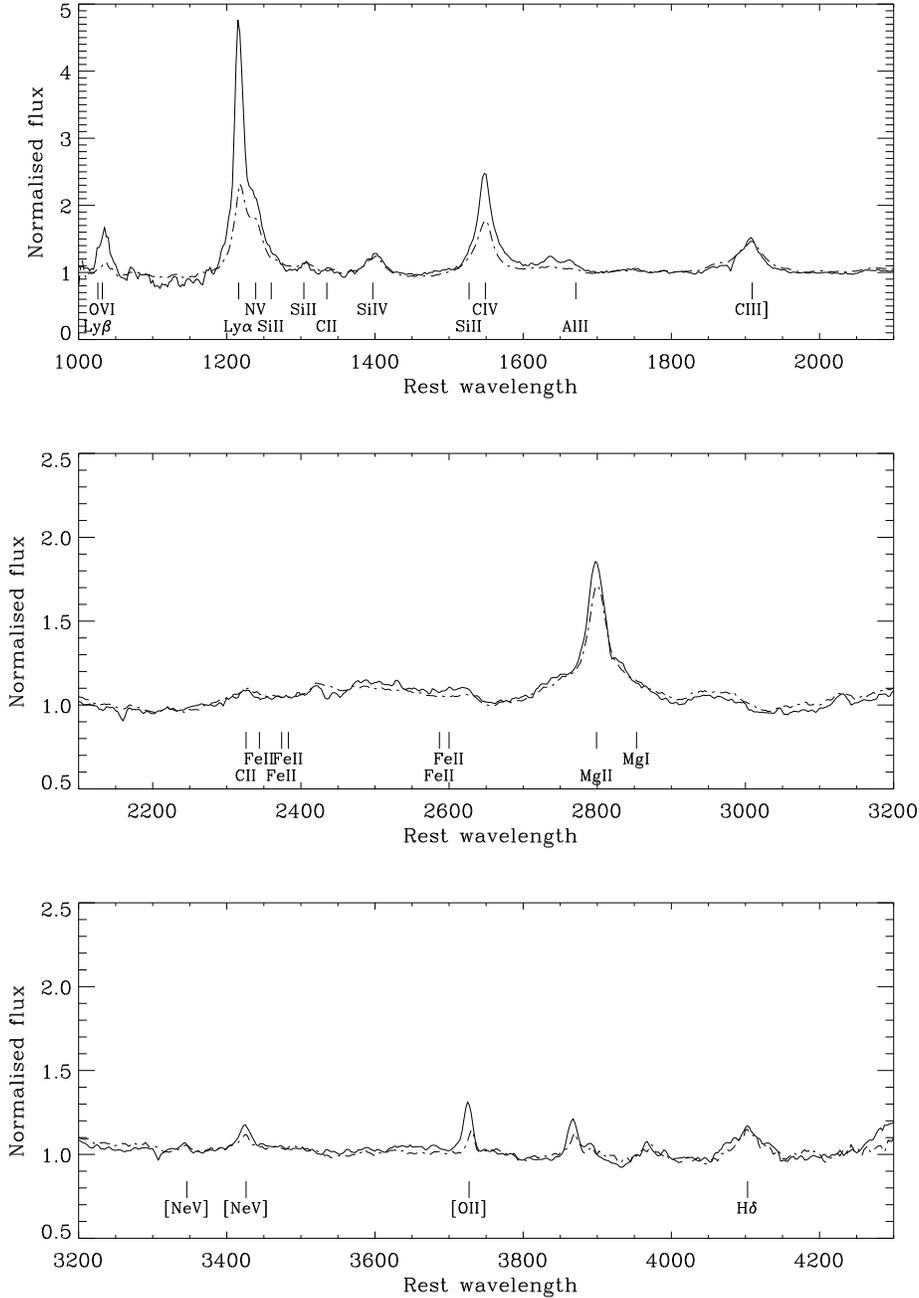,height=6.8in,bbllx=24pt,bblly=23pt,bburx=497pt,bbury=708pt}}
\caption[]{
Comparison of composite spectra for radio-loud and radio-quiet
QSOs.  The dashed line shows the LBQS radio-quiet composite. The
solid line shows the PKS 0.25-0.5Jy (radio loud) quasar composite
described in the text. Both composites have been normalised by their
continua.}
\label{pkscomp}
\end{figure*}

Figure~\ref{pkscomp} shows the comparison of the PKS 0.25-0.5Jy
composite with that for radio-quiet QSOs from the Large Bright
Quasar Survey (LBQS; Hewett, Folz \& Chaffee 1995 and references
therin). This second composite was compiled and provided by P. Francis
(private communication) and is similar to the radio-quiet composite
presented by Francis, Hooper \& Impey (1993) but has been updated to
include a factor $\sim3$ more QSOs, having made use of radio
information from the FIRST survey (Becker, White \& Helfand 1995) to
distinguish between radio-loud and radio quiet QSOs in the LBQS. We
refer to this new composite as the ``LBQS radio-quiet composite.''

The continua of the two composites were found to have somewhat
different slopes, the PKS 0.25-0.5Jy composite having a redder
UV/optical continuum spectral index of $\alpha=-1.0$ compared to
$\alpha=-0.32$ for the LBQS composite ($F(\nu) \propto \nu
^{\alpha}$). However this may be at least in part an artefact caused
by several systematic effects, as described in Francis et al. (1991).
For example, the individual spectra that make up the PKS 0.25-0.5Jy
composite were taken with a fairly narrow slit (typically $1.5''$
wide) and therefore flux may be lost at the ends of the spectra due to
atmospheric dispersion. No attempt was made to correct for
this. Because the composite was built up by including spectra in
redshift sequence, these relative flux calibration errors may have the
effect of producing a gradual gradient in the overall spectral shape
of the composite.

For the purposes of plotting the two composites together, both were
normalised by low-order polynomial fits to their continua (avoiding
regions with strong lines for the continuum fit).  There are some
clear qualitative differences between the composite spectra shown in
Figure~\ref{pkscomp}. The main differences are that the
Ly-$\alpha$(1216\AA) and CIV(1549\AA) lines are stronger in the PKS
0.25-0.5Jy composite.  Below we make a quantitative comparison of the
emission line strengths relative to the local continuum.

\subsection{Rest-frame equivalent width measurements}

We have measured the rest-frame equivalent width (EW) of emission
lines from the PKS 0.25-0.5Jy composite (Figure~\ref{pkscomp}) and
from an electronic version of the LBQS radio-quiet composite. This was
done following the algorithm in Francis (1993) but using slightly
different continuum and line wavelength ranges. The wavelength regions
used were as follows: Ly$\alpha$ line $1216-1230$\AA\ (i.e. the red
side of the line only, since the blue side is affected by Ly$\alpha$
absorption); Ly$\alpha$ continuum $1275-1295$\AA\ and $1440-1480$\AA;
CIV line $1480-1590$\AA; CIV continuum $1450-1480$\AA\ and
$1680-1700$\AA; CIII] line $1850-1940$\AA; CIII] continuum
$1770-1820$\AA\ and $1970-2030$\AA; Fe II line complex $2300-2600$\AA;
FeII continuum $2230-2260$\AA\ and $2650-2670$\AA; MgII line
$2680-2900$\AA; MgII continuum $2640-2670$\AA\ and $3010-3060$\AA.

\begin{figure*}
\centering
\mbox{\psfig{figure=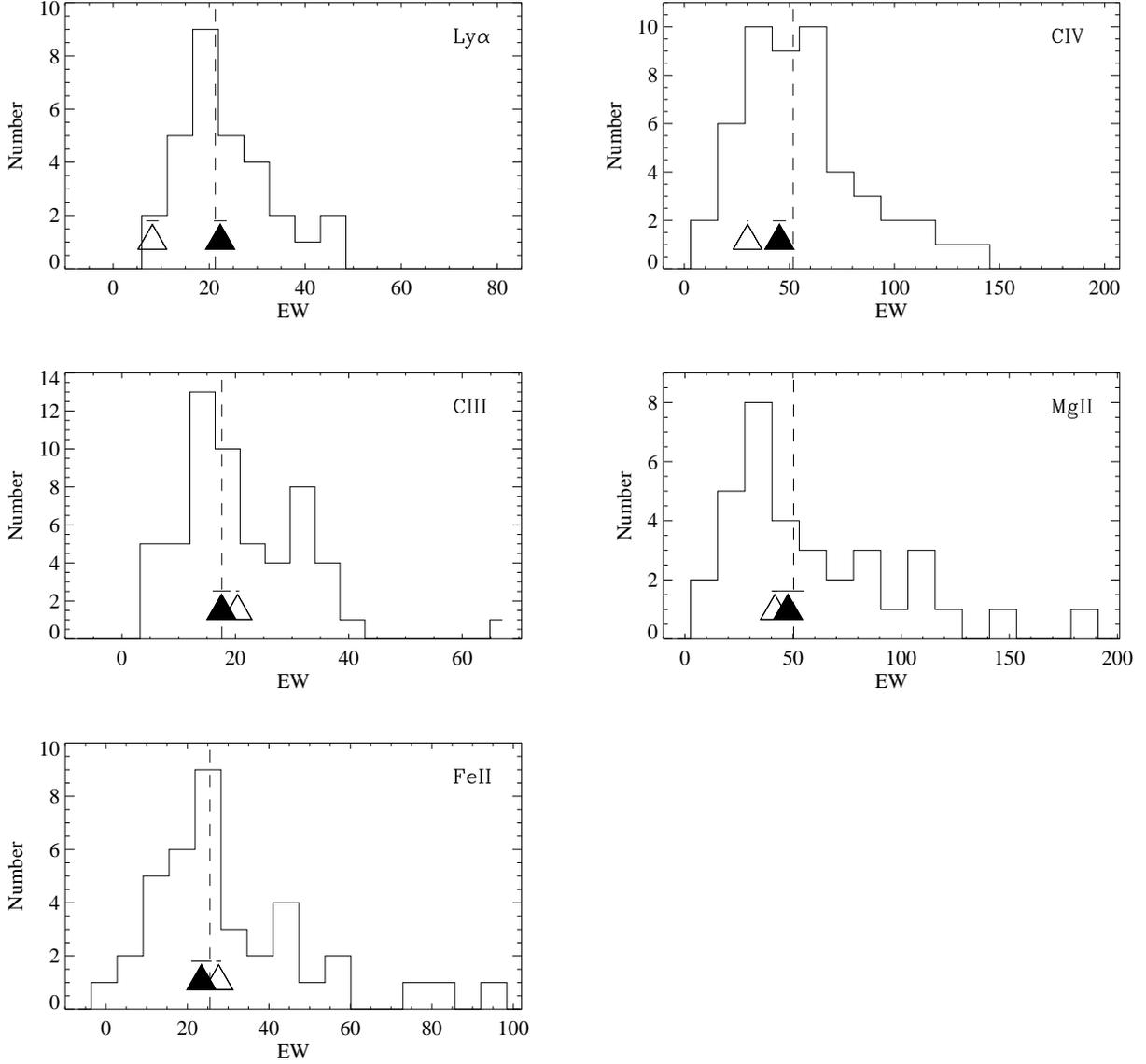,height=6.0in,bbllx=22pt,bblly=7pt,bburx=543pt,bbury=500pt}}
\caption[]{Histograms of rest-frame equivalent widths measured from our PKS 0.25-0.5Jy spectra. Over-plotted are the median of the distribution (dashed line), the 
measurement from the PKS 0.25-0.5Jy composite spectrum (solid
triangle) and the measurements from the LBQS radio-quiet composite
spectrum (open triangle). The assumed error bars for the composite
measurements are plotted as horizontal bars above the triangles.}
\label{ew_dist}
\end{figure*}

The ratios of the EWs are given in Table~\ref{ewtab}. For the PKS
0.25-0.5Jy sample we also measured the EWs of the lines from the
individual spectra, and their distributions are shown in
Figure~\ref{ew_dist}. These distributions were used to estimate the
dispersion of EW measurements. The mean deviation from the median,
divided by $\rm \sqrt{N}$ (where N is the number of measurements of
that line) was taken as an estimate of the uncertainty of the EW
measured from the composite. For the LBQS composite the uncertainty in
the EW measurements was derived using the mean and uncertainties
given in Francis, Hooper \& Impey (1993).  Since those values
correspond to a previous version of their LBQS radio-quiet composite,
we have scaled the uncertainties by the inverse square root of the
relative numbers of QSOs contributing to each line (which typically
reduces the uncertainty by a factor $\sim 2$). The fractional
uncertainty in the EW ratio was taken as the quadratic sum of the
fractional uncertainties determined above.

\begin{table*}

\caption{Rest-frame equivalent widths measured from our data compared with measurements from the literature.
Column (a) gives the number of QSOs contributing to the statistics
for our PKS 0.25-0.5Jy sample.  Column (b) gives the median of the EW
distributions from our PKS 0.25-0.5Jy sample. Column (c) lists the
ratio of EW measurements determined from the PKS 0.25-0.5Jy composite
and the LBQS Radio-quiet composite. The uncertainty
in this estimate is derived from the distribution of EW measurements
from the individual spectra that make up each composite (see text).
Column (d) lists the ratio of EW measurements for radio-loud and
radio-quiet subsets of the LBQS sample from Francis, Hooper \&
Impey (1993).}
\centering
\label{ewtab}
\begin{tabular}{l  c c c c c}\hline
                      & (a)     & (b)            & (c) Composite Ratio      & (d)       \cr 
Line &   N            &  Median EW (\AA) & EW(PKS 0.25-0.5Jy) & EW(loud)/EW(quiet)    \cr 
     &(PKS 0.25-0.5Jy)&(PKS 0.25-0.5Jy)&  /EW(LBQS RQ)       & Francis, Hooper \& Impey (1993)\cr\hline
Ly$\alpha$            &  30     & $21.3\pm1.3$ &  $2.73\pm 0.19$ & $1.56\pm0.34$ \cr
CIV                   &  49     & $51.8\pm3.1$ &  $1.50\pm 0.11$ & $1.47\pm0.24$ \cr
CIII]                 &  57     & $17.6\pm1.6$ &  $0.86\pm 0.08$ & $1.00\pm0.11$ \cr
FeII                  &  36     & $25.5\pm2.5$ &  $0.85\pm 0.10$ & $1.06\pm0.23$ \cr
MgII                  &  36     & $50.3\pm7.6$ &  $1.15\pm 0.18$ & $0.96\pm0.12$ \cr\hline
\end{tabular}
\newline
\end{table*}

\begin{figure}
\centering
\mbox{\psfig{figure=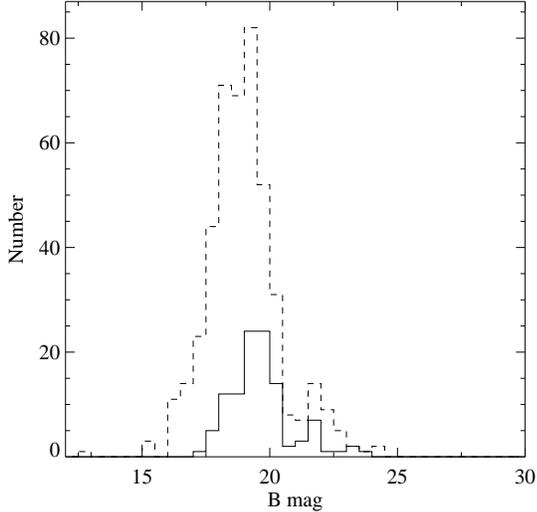,height=2.7in,bbllx=29pt,bblly=14pt,bburx=528pt,bbury=485pt}}\hspace{0.2cm}
\caption[]{Histogram of B magnitudes for all 449 quasars in the Parkes quarter-Jansky
flat-spectrum sample (dashed line) and for the subset of 109 with new
spectra and $\rm 0.25Jy<S_{2.7GHz}< 0.5Jy$ that were used to make the composite
(solid line).}
\label{bmagdist}
\end{figure}

Table~\ref{ewtab} and Figure~\ref{ew_dist} show that the Ly$\alpha$
and CIV lines are significantly stronger in the PKS 0.25-0.5Jy sample
than in the LBQS radio-quiet sample at greater than the
$3\sigma$ level.  This difference is unlikely to be a result of the
Baldwin effect since the optical magnitude distribution of the LBQS,
$m_{\rm B_J}\le18.85$, and the redshift distribution (Hewett, Foltz \&
Chaffee 1995) are similar to those of the sample of quasars used to
make our PKS 0.25-0.5Jy composite (see solid lines in Figures
~\ref{pks_zhist} and ~\ref{bmagdist}).

Previous authors have compared EW distributions of radio-loud and
radio-quiet QSOs but found no significant differences (e.g. Steidel
\& Sargent 1991, Corbin 1992). However, as pointed out by Francis, Hooper \&
Impey (1993), this is generally because of small sample sizes and in
fact these studies are consistent with the measurements of Francis,
Hooper \& Impey (1993) and with those in this paper to within
1$\sigma$. The more recent measurement of the CIV equivalent width
ratio from Corbin \& Francis (1994) of 1.21$\pm$0.21 is also
consistent with our work, although our measurements have lower
uncertainties than these samples individually or combined.

The MgII line, CIII] line and the region dominated by iron lines (FeII)
around 2400\AA-2600\AA\ do not show significant differences in
equivalent width distribution between the PKS radio-loud and LBQS
radio-quiet samples, in agreement with Francis, Hooper \& Impey
(1993).

We have also carried out a similar comparison using the combined LBQS
composite of Francis et al. (1991), which contains a small fraction
($\sim 10\%$) of radio-loud quasars. As would be expected the results
were very similar to those shown in Table~\ref{ewtab} since the
combined LBQS composite is dominated by radio-quiet QSOs. When
considering the ratios of EWs relative to the PKS 0.25-0.5Jy sample,
as in column (c) of Table~\ref{ewtab}, only the ratio of Ly$\alpha$
EWs changed by more than $1\sigma$. Using the combined composite the
ratio was $2.12\pm0.17$ compared to $2.73\pm0.19$.

Although the optical magnitudes of the PKS 0.25-0.5Jy sample and the
LBQS sample are similar, the PKS sample is radio-selected and hence
contains more extreme radio-loud objects than the radio-loud LBQS
sub-sample in Francis, Hooper \& Impey (1993). Almost all the LBQS
sample members, including the radio-loud quasars, have radio flux
densities well below the 0.25Jy flux limit of our sample (Hooper et
al. 1995). As a result we might expect to see more pronounced
differences between the PKS 0.25-0.5Jy composite spectrum and the LBQS
radio-quiet composite than were seen between the radio-loud and radio
quiet subsets of the LBQS sample. However we in fact find very similar
ratios of EWs as were found by Francis, Hooper
\& Impey (1993) for all the lines considered except Ly$\alpha$, which appears to be significantly stronger in the PKS 0.25-0.5Jy sample.

Brotherton et al. (2001) saw no significant difference in the
Ly$\alpha$ line (or indeed any of the UV lines that we have studied)
when they considered composite spectra of radio-loud and radio-quiet
objects from the FIRST Bright Quasar Survey (FBQS). That sample is
radio-selected but the faint flux density limit ($\rm S_{1.4GHz}\sim
1mJy$) and bright optical limit ($\rm E<17.8$) means it is still
sensitive to quasars that are formally radio-quiet, although as they
point out, the extreme radio-quiet and ``radio-silent'' QSOs are not
represented.  It is possible that their result, that of Francis,
Hooper \& Impey (1993) and our result represent a trend of Ly$\alpha$
EW becoming progressively stronger with increased radio
loudness. However, as Brotherton et al. (2001) also point out, their
radio-quiet sub-sample has rather limited statistics - only $\sim 12$
quasars contribute to the Ly$\alpha$ line in their radio-quiet
composite. Because of the difference in optical magnitude
distributions of the FBQS and our sample, and hence the possible
complication of the Baldwin effect, we have not carried out a detailed
comparison with the FBQS composites.

We also note that the EWs of Ly$\alpha$
and CIV in classical radio galaxies are comparable but somewhat
stronger than those found for the PKS 0.25-0.5Jy sample: values of
EW=919\AA\ for Ly$\alpha$ and 74\AA\ for CIV are reported by McCarthy
(1993), measured from a composite of high-redshift 3CR and 1Jy class
radio galaxies. This, along with the appearance of HeII1640 and
stronger forbidden lines (see Figure 5), puts the PKS 0.25-0.5Jy
sources on a smooth trend from radio-quiet to flat-spectrum quasar to
steep-spectrum quasar to radio galaxy. This trend, at least among
radio-loud quasars, has been interpreted by previous authors in the
context of unified schemes of AGN. For example Baker \& Hunstead
(1995) found that the EW of narrow lines in radio-loud quasars from
the Molonglo sample increases as the ratio of core-to-lobe flux (R)
decreases, i.e. as the implied viewing angle to the radio jet axis
increases.

The EW ratios from composite spectra given in Table~\ref{ewtab} are
relative measurements using the same method on each composite and are
therefore largely independent of the details of the continuum fitting
and flux integration method used to determine the EWs.  The absolute
EW measurements shown in Figure~\ref{ew_dist} were only used to
estimate the uncertainty in the value measured from the PKS 0.25-0.5Jy
composite.

We note that Francis et al. (2002) have measured rest-frame EWs from
the stronger, flat spectrum radio sources ($\rm S_{2.7GHz}\ge 0.5Jy$)
in the PKS sample. We have used the the electronic table from that
paper to derive median EWs of $40.13\pm3.40$, $20.61\pm1.75$ and
$29.61\pm1.13$ for CIV, CIII] and MgII respectively (this combines the
data from their red and blue sub-samples, which showed no significant
differences in these lines). The value for CIII] is similar to ours for
weaker PKS 0.25-0.5Jy radio sources, but the median values for CIV and
MgII are lower than ours at the $2-3\sigma$ level and closer to the
values we derive for the LBQS composite. However, since this comparison
involves absolute EW measurements made by different groups and hence
may be affected by systematic measurement effects, we do not consider
the differences to be significant.

\section{The Lyman-$\alpha$ flux decrement as a function of Redshift}

Our spectra can also be used to measure the mean flux decrement
shortward of the Ly$\alpha$ emission line as a function of redshift.
We used the same method as previous authors (e.g. Oke \& Korycansky
1982, Schneider et al. 1991b), namely to calculate
\begin{equation}
       D_A=  1 -  {f(observed)\over{f(continuum)}}
\end{equation}
where f(observed) is the observed flux averaged over the rest-frame
region $1050-1170$\AA\ and f(continuum) is the expected continuum flux
in the same region, which is estimated by extrapolating the continuum
redward of the Ly$\alpha$ emission line into the blue.

\begin{figure}
\centering
\mbox{\psfig{figure=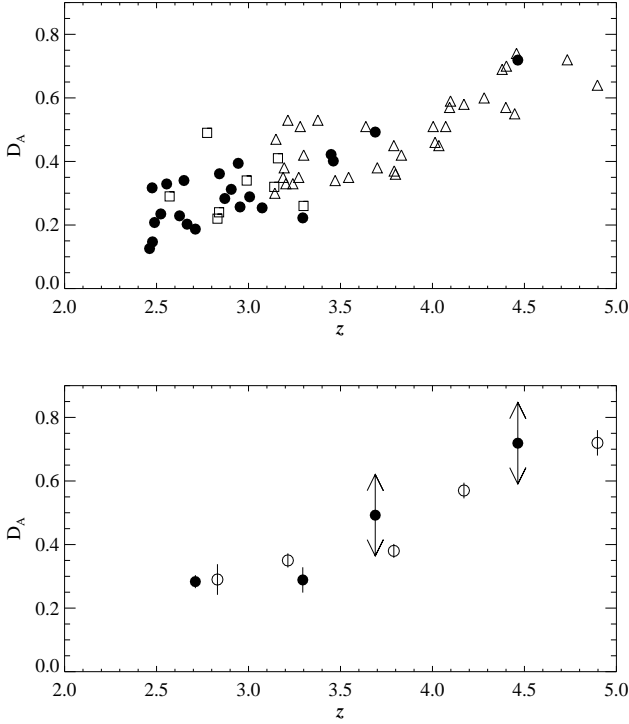,height=3.8in,bbllx=78pt,bblly=245pt,bburx=468pt,bbury=694pt}}
\caption[]{Upper panel: The Lyman-$\alpha$ decrement $D_A$ as a function of redshift.
Filled circles represent quasars from the Parkes quarter-Jansky
flat-spectrum sample, open symbols represent optically-selected QSOs
from the literature. The latter points are taken from Schneider,
Schmidt and Gunn (1991a, 1991b; triangles) and from Oke \& Korycansky
(1982) and Steidel \& Sargent (1987; squares). Lower panel: The same
data in bins of 0.5 in redshift, with the optically-selected samples
combined. The points are plotted at the median ($z, D_A$) within each
bin. 1-$\sigma$ error bars are plotted except for cases where the bin
only contains one object in which case arrows are drawn.}
\label{daz}
\end{figure}

There were 22 quasars in our sample for which we obtained a spectrum
and which have sufficiently high redshift ($z \ge 2.4 $) for the
region around the Ly-$\alpha$ line to be covered.  (Note that the full
1050-1170\AA\ region is not covered in all cases but enough of the
region is covered to give a good measurement of $D_A$.)  The $D_A$
values are plotted against redshift for the PKS QSOs in
Figure~\ref{daz}.  For comparison, values are plotted for optically
selected QSOs taken from Schneider, Schmidt and Gunn (1991a, 1991b).
Also plotted are other values from the literature (Oke \& Korycansky
1982, Steidel \& Sargent 1987). Only optically-selected QSOs from
these papers are plotted.

In the lower panel of Figure~\ref{daz} the same data are plotted in
bins of 0.5 in redshift. Points are plotted at the median redshift and
$D_A$ values within each bin. The errors in the median were estimated
by taking the standard deviation in each bin and dividing by
$\sqrt{N}$ (the bins containing only one point have undefined error
bars). The plot shows that there is no significant difference between
the two sets of points. This indicates that neither sample is
significantly biased towards or against QSOs with a large amount of
Ly-$\alpha$ absorption.

\section{Conclusions}

We have presented spectra and redshift measurements for a large sample
of radio-selected quasars which, when combined with data from the
literature, forms a complete sample of radio-loud quasars. We have
studied statistical properties of the new spectra, in particular the
rest-frame equivalent width distribution of the rest-frame UV spectral
lines and the Lyman-$\alpha$ decrement at high redshift. We find that
the spectra we have obtained show significantly stronger Ly$\alpha$
and CIV emission lines than radio-quiet quasars from the LBQS, which
has a similar optical magnitude distribution to our sample. We find no
major difference in the Lyman-$\alpha$ decrement measured from our
spectra compared to those of optically-selected QSOs at similar
redshifts.

\acknowledgements

We are grateful to Paul Francis for generating and making available a
new version of the composite spectrum of LBQS radio-quiet QSOs.  We
also thank the helpful support staff at the ESO 3.6m, where the
majority of our spectroscopic observations were carried out.  This
paper has benefited from helpful comments from the referee,
Dr. P. J. P. McCarthy.

\begin{figure*}
\centering
\mbox{\psfig{figure=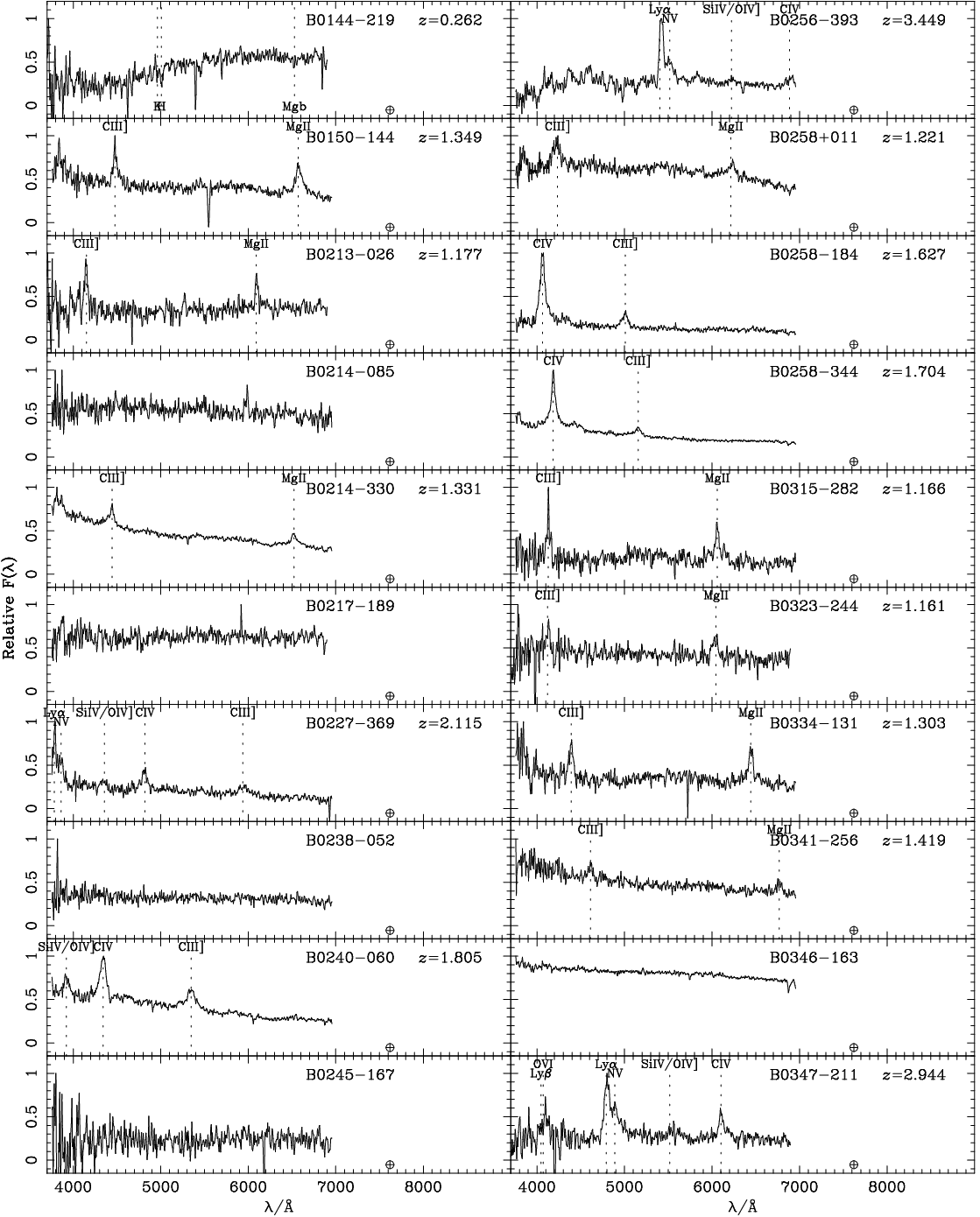,width=7.0in,bbllx=48pt,bblly=60pt,bburx=587pt,bbury=732pt}}
{\bf Fig. 1.} {\it continued}
\end{figure*}
\begin{figure*}
\centering
\mbox{\psfig{figure=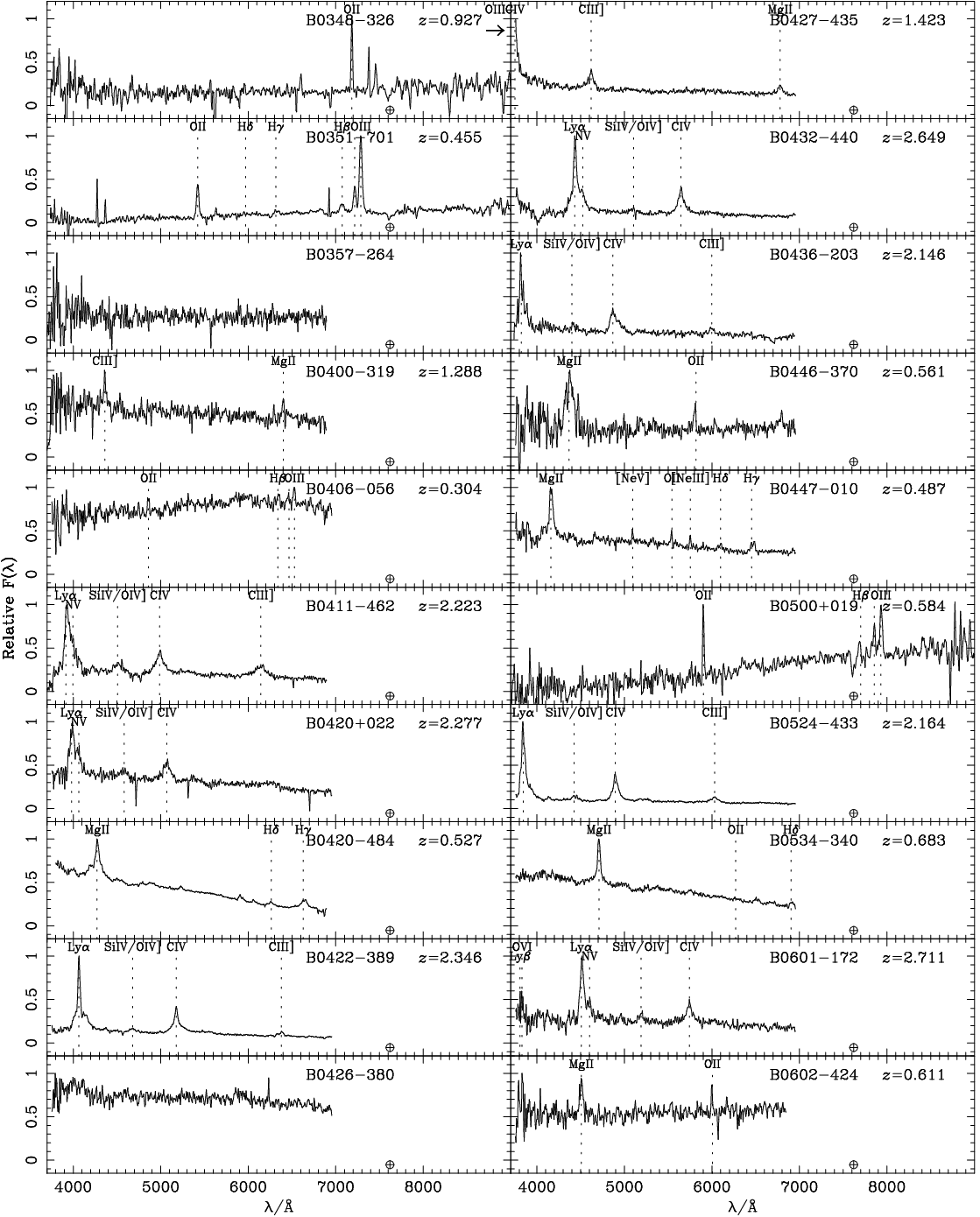,width=7.0in,bbllx=48pt,bblly=60pt,bburx=587pt,bbury=732pt}}
{\bf Fig. 1.} {\it continued}
\end{figure*}
\begin{figure*}
\centering
\mbox{\psfig{figure=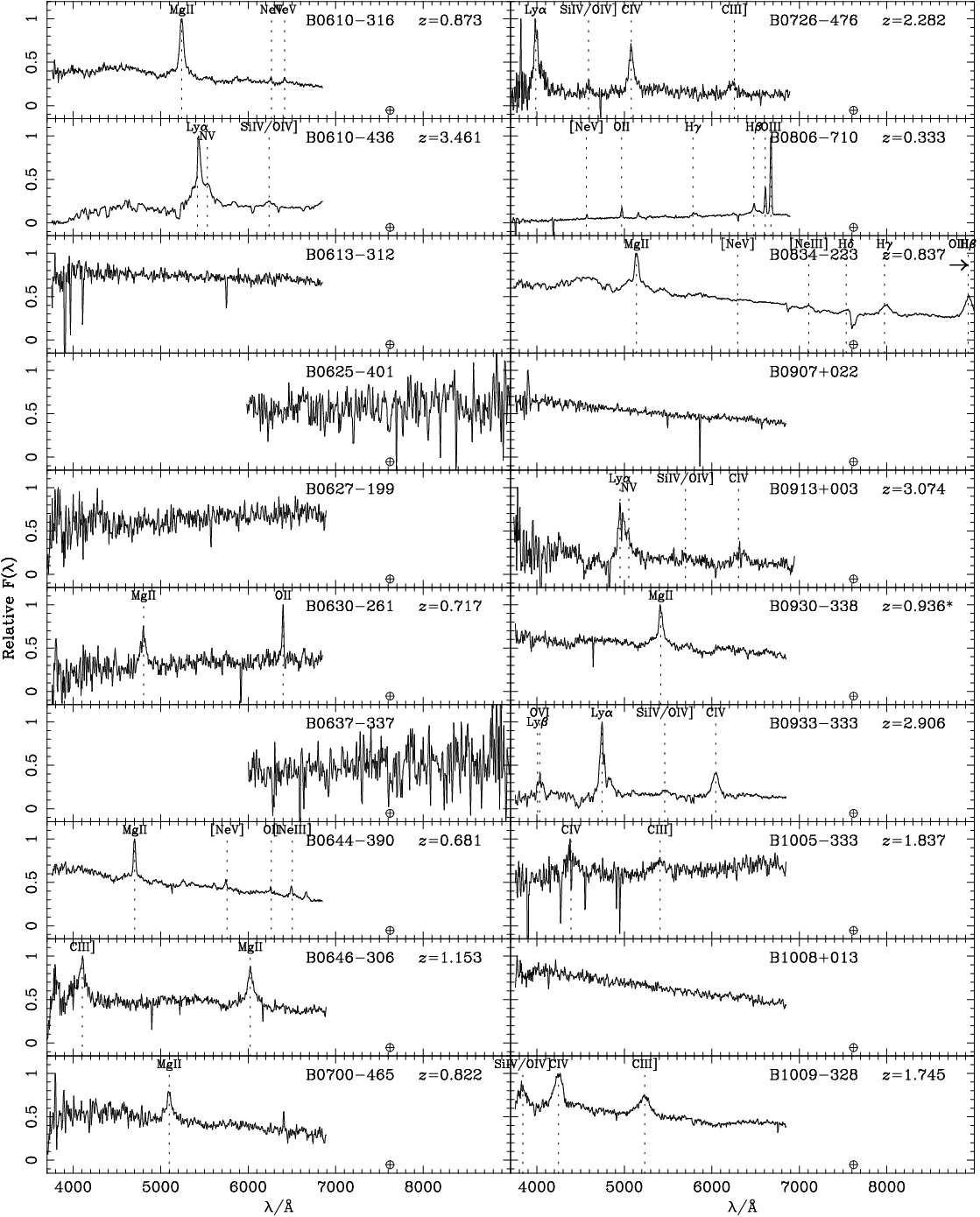,width=7.0in,bbllx=48pt,bblly=60pt,bburx=587pt,bbury=732pt}}
{\bf Fig. 1.} {\it continued}
\end{figure*}
\begin{figure*}
\centering
\mbox{\psfig{figure=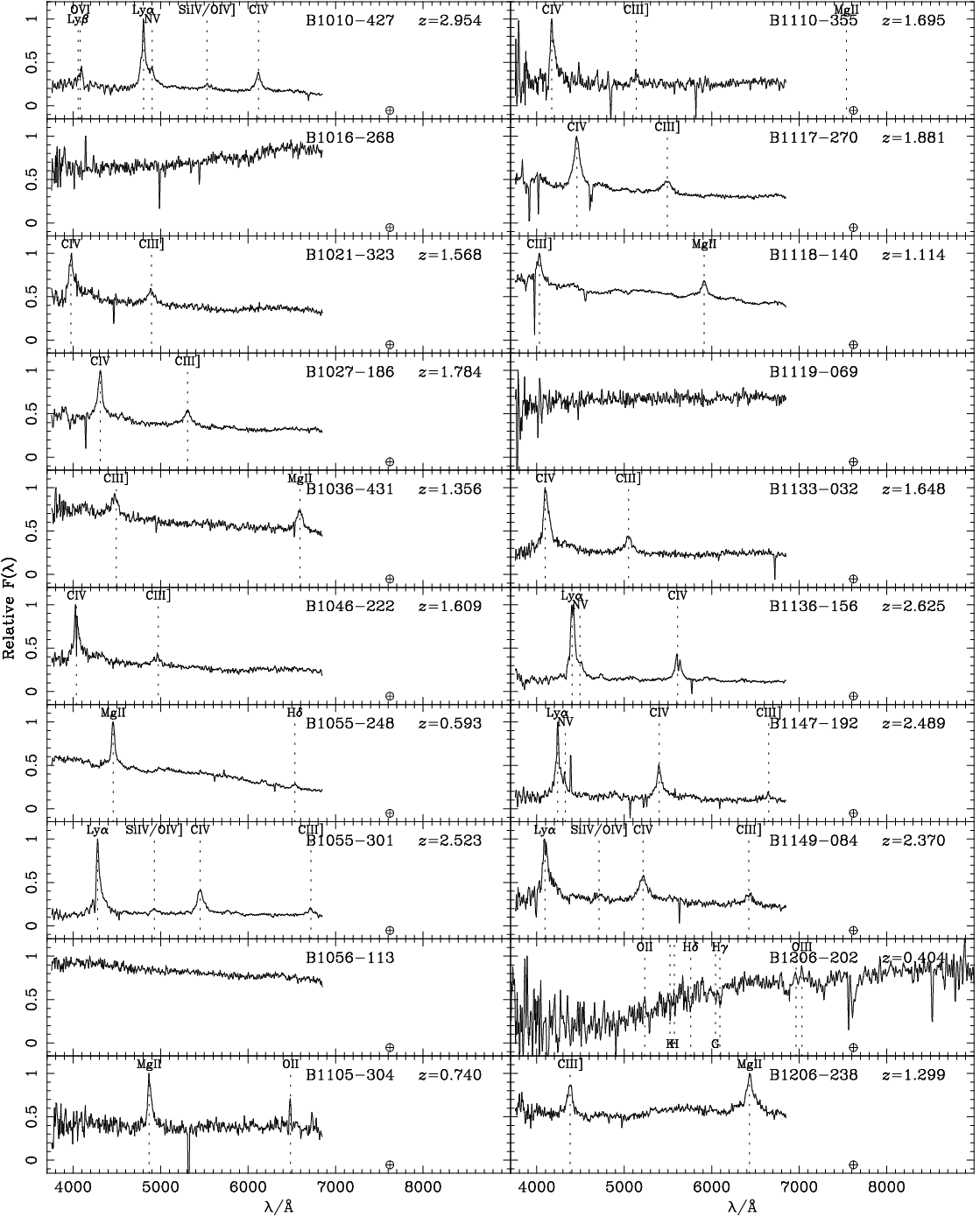,width=7.0in,bbllx=48pt,bblly=60pt,bburx=587pt,bbury=732pt}}
{\bf Fig. 1.} {\it continued}
\end{figure*}
\begin{figure*}
\centering
\mbox{\psfig{figure=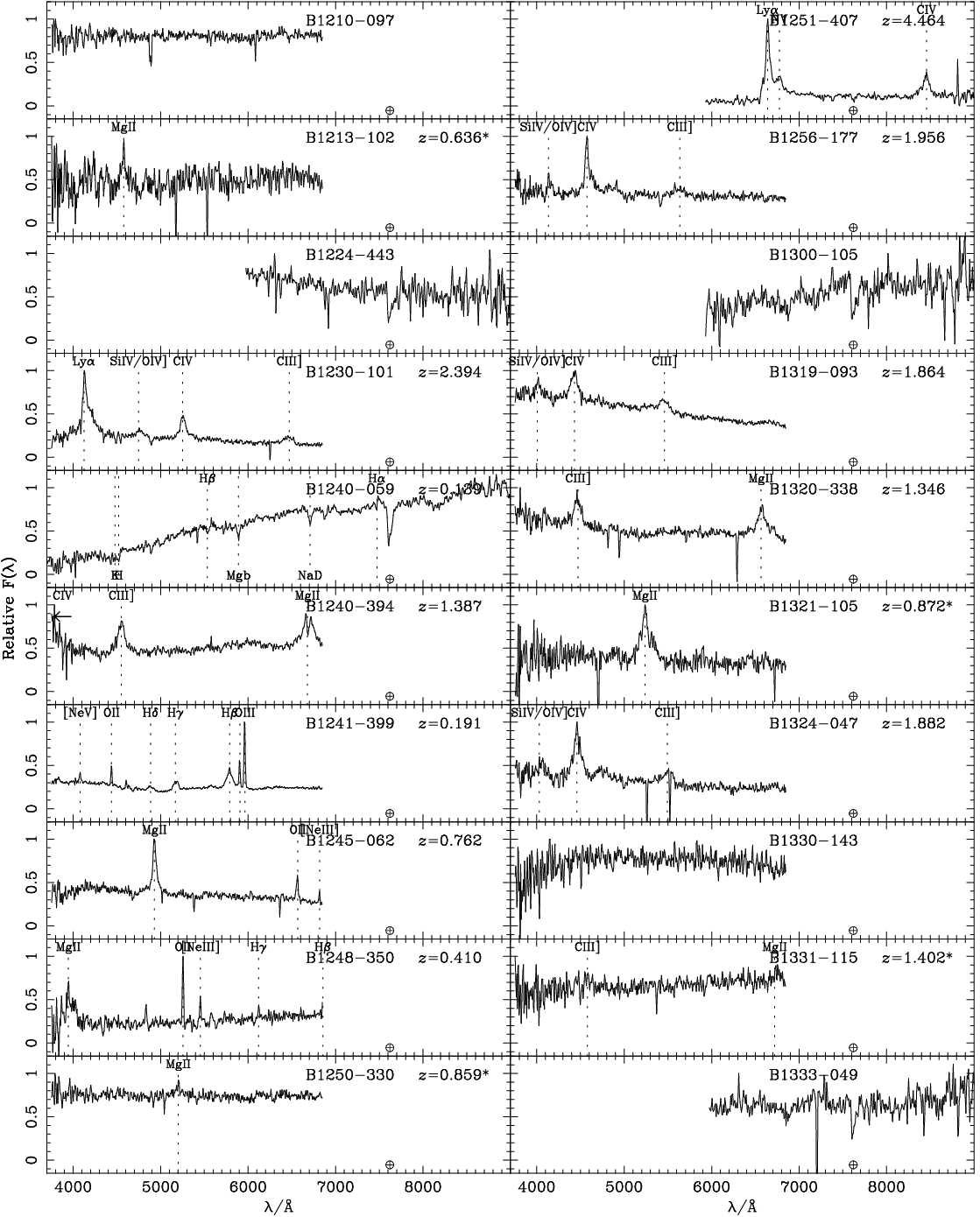,width=7.0in,bbllx=48pt,bblly=60pt,bburx=587pt,bbury=732pt}}
{\bf Fig. 1.} {\it continued}
\end{figure*}
\begin{figure*}
\centering
\mbox{\psfig{figure=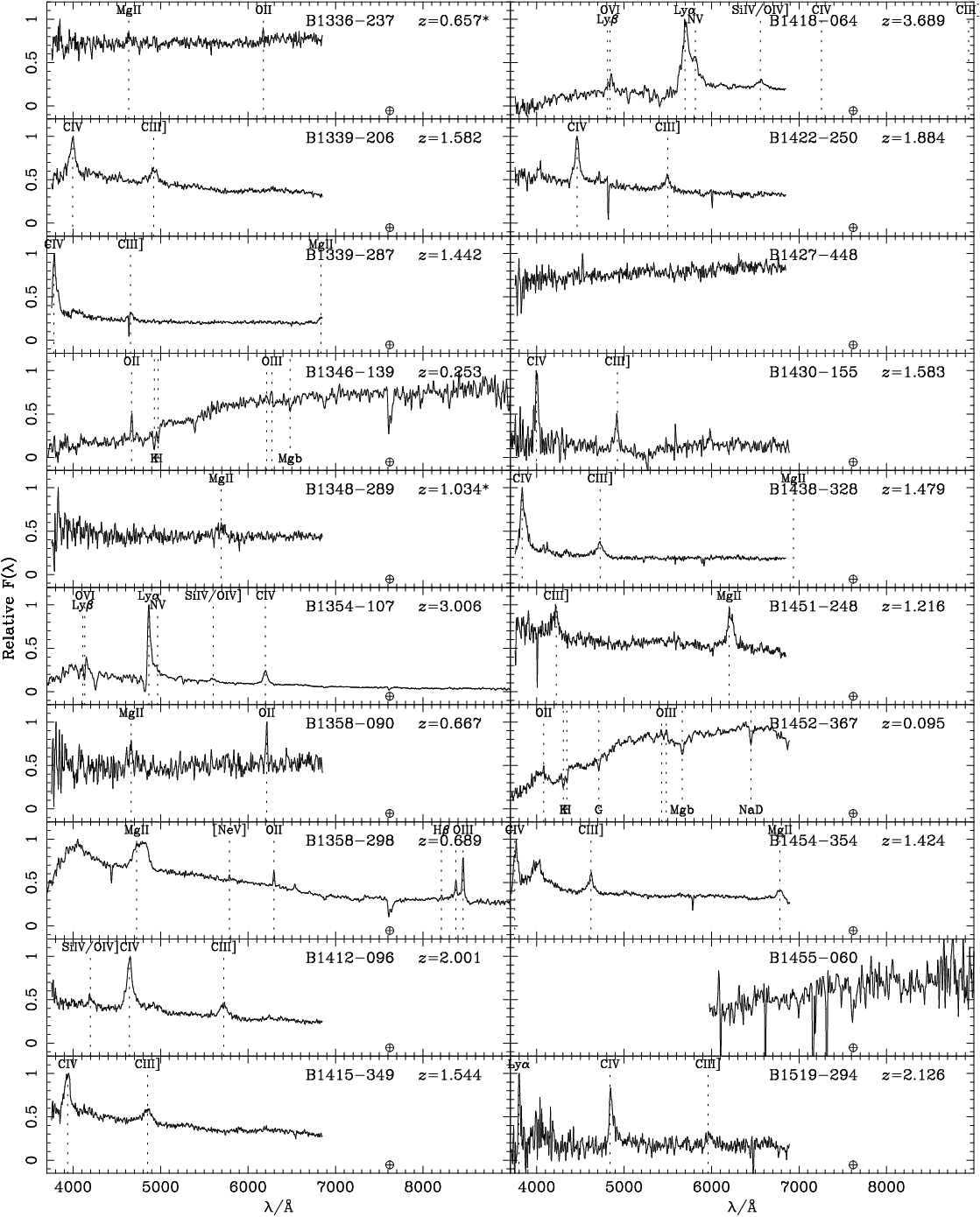,width=7.0in,bbllx=48pt,bblly=60pt,bburx=587pt,bbury=732pt}}
{\bf Fig. 1.} {\it continued}
\end{figure*}
\begin{figure*}
\centering
\mbox{\psfig{figure=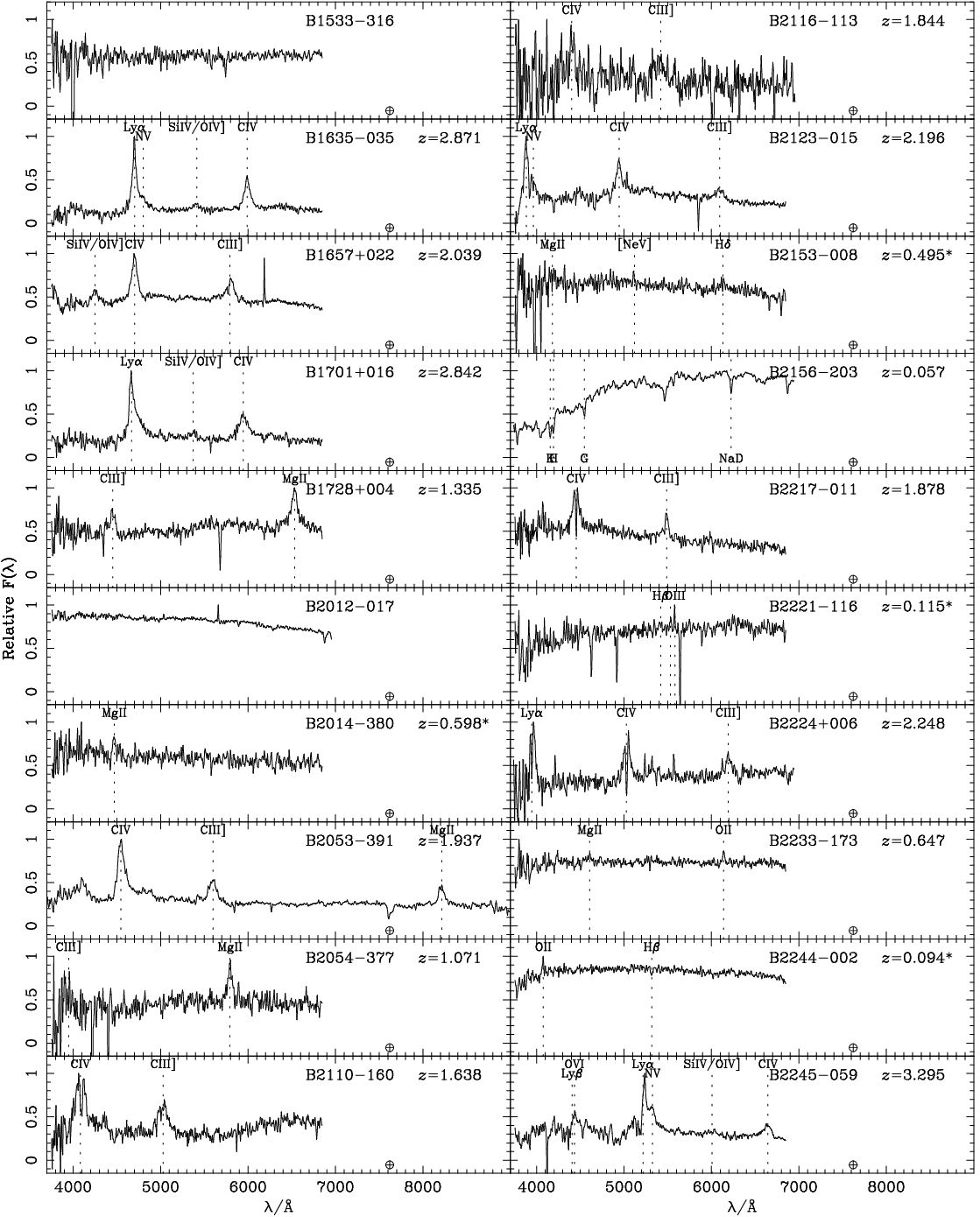,width=7.0in,bbllx=48pt,bblly=60pt,bburx=587pt,bbury=732pt}}
{\bf Fig. 1.} {\it continued}
\end{figure*}
\begin{figure*}
\centering
\mbox{\psfig{figure=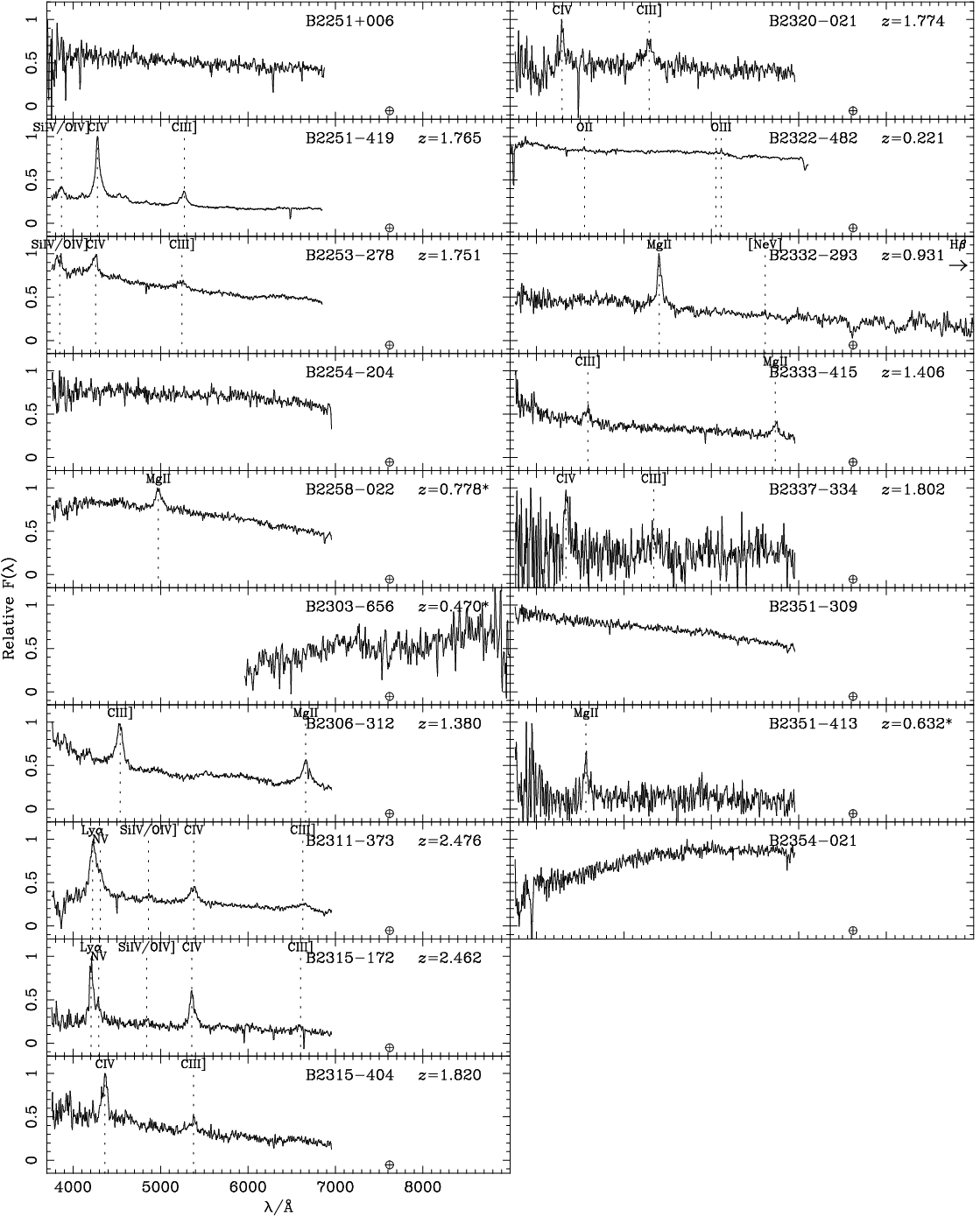,width=7.0in,bbllx=48pt,bblly=60pt,bburx=587pt,bbury=732pt}}
{\bf Fig. 1.} {\it continued}
\end{figure*}

\appendix
\section{Tables}

\begin{table}
\caption[]{Redshifts for the Parkes quarter-Jansky flat-spectrum
sample. An asterisk next to the name indicates a note in the text for
that object. The ``conf'' column gives a level of confidence in the
redshift (0=no lines for redshift determination, 1=secure redshift,
2=single line or weak lines).  The Class column gives the
classification, reproduced from Jackson et al. (2002). These
classifications are based on both the spectrum and images of the
source and their meanings are as follows: Q =confirmed quasar by
spectroscopy or UVX photometry; BL = spectroscopically-confirmed BL
Lac object; G = Galaxy with morphological or spectroscopic
confirmation; NCS = Non-compact radio source (the source is either a
steep-spectrum or extended radio source). Refer to individual notes in
Jackson et al (2002) for full details.  }
\begin{tabular}{lcrll}\hline
Name & $z$ & Conf& Class & Date\cr\hline
B0000$-$160& 0.508& 1&G   & Oct 95           \cr
B0008$-$307& 1.189& 1&Q   & Oct 97                           \cr
B0012$-$184& 0.743& 1&Q   & Oct 97                           \cr
B0017$-$307& 2.666& 1&Q   & Oct 97                           \cr
B0022$-$044& 1.946& 1&Q   & Oct 97                           \cr
B0026$-$014& 0.083& 1&G   & Oct 95                           \cr
B0027$-$426& 0.495& 1&Q   & Oct 97                           \cr
B0038$-$326& $-$  & 0&BL  & Oct 97                           \cr
B0039$-$407$^*$& 2.478& 1&Q   & Oct 97                       \cr
B0056$-$525& $-$  & 0&BL  & Oct 94                           \cr
B0059$-$287& 1.694& 1&Q   & Oct 95                           \cr
B0111$-$256& 1.030& 1&Q   & Oct 97                           \cr
B0113$-$283& 2.555& 1&Q   & Oct 97                           \cr
B0115$-$016& 1.162& 1&Q   & Oct 95                           \cr
B0115$-$342& 0.647& 1&Q   & Oct 97                           \cr
B0118$-$272& $-$  & 0&BL  & Oct 97                           \cr
B0136$-$059& 2.004& 1&Q   & Oct 97                           \cr
B0138$-$097$^*$& $-$  & 0&BL  & Oct 97                           \cr
B0140$-$059& $-$  & 0&BL  & Oct 97                           \cr
B0143$-$061& 0.499& 1&NCS & Oct 95                           \cr
B0144$-$219& 0.262& 1&G   & Oct 94                           \cr
B0150$-$144& 1.349& 1&Q   & Oct 97                           \cr
B0213$-$026& 1.177& 1&Q   & Oct 94                           \cr
B0214$-$085& $-$  & 0&BL  & Oct 97                           \cr
B0214$-$330& 1.331& 1&Q   & Oct 97                           \cr
B0217$-$189& $-$  & 0&BL  & Oct 97                           \cr
B0227$-$369& 2.115& 1&Q   & Oct 97                           \cr
B0238$-$052& $-$  & 0&BL  & Oct 97                           \cr
B0240$-$060& 1.805& 1&Q   & Oct 97                           \cr
B0245$-$167& $-$  & 0&BL  & Oct 97                           \cr
B0256$-$393& 3.449& 1&Q   & Oct 97                           \cr
B0258$+$011& 1.221& 1&Q   & Oct 97                           \cr
B0258$-$184& 1.627& 1&Q   & Oct 97                           \cr
B0258$-$344& 1.704& 1&Q   & Oct 97                           \cr
B0315$-$282& 1.166& 1&Q   & Oct 97                           \cr
\end{tabular}
\label{pksz}
\end{table}

\begin{table}
\begin{tabular}{lcrll}\hline
Name & $z$ & Conf& Class & Date \cr\hline
B0323$-$244& 1.161& 1&Q   & Nov 93                           \cr
B0334$-$131& 1.303& 1&Q   & Oct 97                           \cr
B0341$-$256& 1.419& 1&Q   & Oct 97                           \cr
B0346$-$163& $-$  & 0&BL  & Oct 97                           \cr
B0347$-$211& 2.944& 1&Q   & Nov 93                           \cr
B0348$-$326& 0.927& 1&G   & Oct 95                           \cr
B0351$-$701& 0.455& 1&Q   & Oct 95                           \cr
B0357$-$264$^*$& $-$ & 0&BL  & Nov 93                         \cr
B0400$-$319& 1.288& 1&Q   & Nov 93                           \cr
B0406$-$056& 0.304& 1&G   & Oct 97                           \cr
B0411$-$462& 2.223& 1&Q   & Nov 93                           \cr
B0420$+$022& 2.277& 1&Q   & Oct 97                           \cr
B0420$-$484& 0.527& 1&Q   & Oct 95                           \cr
B0422$-$389& 2.346& 1&Q   & Oct 97                           \cr
B0426$-$380& $-$  & 0&BL  & Oct 97                           \cr
B0427$-$435& 1.423& 1&Q   & Oct 97                           \cr
B0432$-$440& 2.649& 1&Q   & Oct 97                           \cr
B0436$-$203& 2.146& 1&Q   & Oct 95                           \cr
B0446$-$370& 0.561& 1&Q   & Oct 97                           \cr
B0447$-$010& 0.487& 1&Q   & Oct 97                           \cr
B0500$+$019& 0.584& 1&G   & Oct 94                           \cr
B0524$-$433& 2.164& 1&Q   & Oct 97                           \cr
B0534$-$340& 0.683& 1&Q   & Oct 97                           \cr
B0601$-$172& 2.711& 1&Q   & Oct 97                           \cr
B0602$-$424& 0.611& 1&Q   & May 97                           \cr
B0610$-$316& 0.873& 1&Q   & May 97                           \cr
B0610$-$436& 3.461& 1&Q   & May 97                           \cr
B0613$-$312& $-$  & 0&BL  & May 97                           \cr
B0625$-$401& $-$  & 0&G   & Oct 95                           \cr
B0627$-$199& $-$  & 0&BL  & Oct 94                           \cr
B0630$-$261& 0.717& 1&Q   & May 97                           \cr
B0637$-$337& $-$  & 0&BL  & May 97                           \cr
B0644$-$390& 0.681& 1&Q   & May 97                           \cr
B0646$-$306$^*$& 1.153& 1&Q   & Nov 93                       \cr
B0700$-$465& 0.822& 1&Q   & Nov 93                           \cr
B0726$-$476& 2.282& 1&Q   & Nov 93                           \cr
B0806$-$710& 0.333& 1&Q   & Apr 95                           \cr
B0834$-$223& 0.837& 1&Q   & Oct 95                           \cr
B0907$+$022& $-$  & 0&BL  & May 97                           \cr
B0913$+$003& 3.074& 1&Q   & Oct 95                           \cr
B0930$-$338& 0.936& 1&Q   & May 97                           \cr
B0933$-$333& 2.906& 1&Q   & May 97                           \cr
B1005$-$333& 1.837& 1&Q   & May 97                           \cr
B1008$+$013& $-$  & 0&BL  & May 97                           \cr
B1009$-$328& 1.745& 1&Q   & May 97                           \cr
B1010$-$427& 2.954& 1&Q   & May 97                           \cr
B1016$-$268& $-$  & 0&BL  & May 97                           \cr
\end{tabular}
\end{table}

\begin{table}
\begin{tabular}{lcrll}\hline
Name & $z$ & Conf& Class & Date\cr\hline
B1021$-$323& 1.568& 1&Q   & May 97                           \cr
B1027$-$186& 1.784& 1&Q   & May 97                           \cr
B1036$-$431& 1.356& 1&Q   & May 97                           \cr
B1046$-$222& 1.609& 1&Q   & May 97                           \cr
B1055$-$248& 0.593& 1&Q   & May 97                           \cr
B1055$-$301& 2.523& 1&Q   & May 97                           \cr
B1056$-$113& $-$  & 0&BL  & May 97                           \cr
B1105$-$304& 0.740& 1&Q   & May 97                           \cr
B1110$-$355& 1.695& 1&Q   & May 97                           \cr
B1117$-$270& 1.881& 1&Q   & May 97                           \cr
B1118$-$140& 1.114& 1&Q   & May 97                           \cr
B1119$-$069& $-$  & 0&BL  & May 97                           \cr
B1133$-$032& 1.648& 1&Q   & May 97                           \cr
B1136$-$156$^*$& 2.625& 1&Q   & May 97                           \cr
B1147$-$192& 2.489& 1&Q   & May 97                           \cr
B1149$-$084& 2.370& 1&Q   & May 97                              \cr
B1206$-$202& 0.404& 1&G   & Apr 95                              \cr
B1206$-$238& 1.299& 1&Q   & May 97                              \cr
B1210$-$097& $-$  & 0&BL  & May 97                              \cr
B1213$-$102& 0.636& 2&Q   & May 97                              \cr
B1224$-$443& $-$  & 0&BL  & Apr 95                              \cr
B1230$-$101& 2.394& 1&Q   & May 97                              \cr
B1240$-$059& 0.139& 1&G   & Apr 95                              \cr
B1240$-$394& 1.387& 1&Q   & May 97                              \cr
B1241$-$399& 0.191& 1&Q   & May 97                              \cr
B1245$-$062& 0.762& 1&Q   & May 97                              \cr
B1248$-$350& 0.410& 1&Q   & May 97                              \cr
B1250$-$330& 0.859& 2&Q   & May 97                              \cr
B1251$-$407$^*$& 4.464& 1&Q   & May 94                              \cr
B1256$-$177& 1.956& 1&Q   & May 97                              \cr
B1300$-$105& $-$  & 0&BL  & May 94                              \cr
B1319$-$093& 1.864& 1&Q   & May 97                              \cr
B1320$-$338& 1.346& 1&Q   & May 97                              \cr
B1321$-$105& 0.872& 2&Q   & May 97                              \cr
B1324$-$047& 1.882& 1&Q   & May 97                              \cr
B1330$-$143& $-$  & 0&BL  & May 97                              \cr
B1331$-$115& 1.402& 2&Q   & May 97                              \cr
B1333$-$049& $-$  & 0&BL  & Apr 95                              \cr
B1336$-$237& 0.657& 2&Q   & May 97                              \cr
B1339$-$206& 1.582& 1&Q   & May 97                              \cr
B1339$-$287& 1.442& 1&Q   & May 97                              \cr
B1346$-$139& 0.253& 1&G   & Apr 95                              \cr
B1348$-$289& 1.034& 2&Q   & May 97                              \cr
B1354$-$107& 3.006& 1&Q   & Apr 95                              \cr
B1358$-$090& 0.667& 1&Q   & May 97                              \cr
B1358$-$298& 0.689& 1&Q   & Apr 95                              \cr
B1412$-$096& 2.001& 1&Q   & May 97                              \cr
B1415$-$349& 1.544& 1&Q   & May 97                              \cr
\end{tabular}
\end{table}

\begin{table}
\begin{tabular}{lcrll}\hline
Name & $z$ & Conf& Class & Date\cr\hline
B1418$-$064& 3.689& 1&Q   & May 97                              \cr
B1422$-$250& 1.884& 1&Q   & May 97                              \cr
B1427$-$448& $-$  & 0&BL  & May 97                              \cr
B1430$-$155& 1.583& 1&Q   & Apr 95                              \cr
B1438$-$328& 1.479& 1&Q   & May 97                              \cr
B1451$-$248& 1.216& 1&Q   & May 97                              \cr
B1452$-$367& 0.095& 1&G   & Apr 95                              \cr
B1454$-$354& 1.424& 1&Q   & Apr 95                              \cr
B1455$-$060& $-$  & 0&BL  & Apr 95                              \cr
B1519$-$294& 2.126& 1&Q   & Apr 95                              \cr
B1533$-$316& $-$  & 0&BL  & May 97                              \cr
B1635$-$035& 2.871& 1&Q   & May 97                              \cr
B1657$+$022& 2.039& 1&Q   & May 97                              \cr
B1701$+$016& 2.842& 1&Q   & May 97                              \cr
B1728$+$004& 1.335& 1&Q   & May 97                           \cr
B2012$-$017& $-$  & 0&BL  & Oct 97                           \cr
B2014$-$380& 0.598& 2&BL  & May 97                           \cr
B2053$-$391& 1.937& 1&Q   & Apr 95                           \cr
B2054$-$377& 1.071& 1&Q   & May 97                           \cr
B2110$-$160& 1.638& 1&Q   & May 97                           \cr
B2116$-$113& 1.844& 1&Q   & Oct 97                           \cr
B2123$-$015& 2.196& 1&Q   & May 97                           \cr
B2153$-$008& 0.495& 2&BL  & May 97                           \cr
B2156$-$203& 0.057& 1&NCS & Oct 95                           \cr
B2217$-$011& 1.878& 1&Q   & May 97                           \cr
B2221$-$116& 0.115& 2&BL  & May 97                           \cr
B2224$+$006& 2.248& 1&Q   & Oct 95                           \cr
B2233$-$173& 0.647& 1&Q   & May 97                           \cr
B2244$-$002& 0.094& 2&G   & May 97                           \cr
B2245$-$059& 3.295& 1&Q   & May 97                           \cr
B2251$+$006& $-$  & 0&BL  & Oct 94                           \cr
B2251$-$419& 1.765& 1&Q   & May 97                           \cr
B2253$-$278& 1.751& 1&Q   & May 97                           \cr
B2254$-$204& $-$  & 0&BL  & Oct 97                           \cr
B2258$-$022& 0.778& 2&Q   & Oct 97                           \cr
B2303$-$656$^*$& 0.47 & 2&G   & Oct 94                       \cr
B2306$-$312& 1.380& 1&Q   & Oct 97                           \cr
B2311$-$373& 2.476& 1&Q   & Oct 97                           \cr
B2315$-$172& 2.462& 1&Q   & Oct 97                           \cr
B2315$-$404& 1.820& 1&Q   & Oct 97                           \cr
B2320$-$021& 1.774& 1&Q   & Oct 97                           \cr
B2322$-$482& 0.221& 1&BL  & Nov 93                           \cr
B2332$-$293& 0.931& 1&Q   & Oct 97                           \cr
B2333$-$415& 1.406& 1&Q   & Oct 97                           \cr
B2337$-$334& 1.802& 1&Q   & Oct 97                           \cr
B2351$-$309& $-$  & 0&BL  & Oct 97                           \cr
B2351$-$413& 0.632& 2&Q   & Oct 97                           \cr
B2354$-$021& $-$  & 0&BL  & Oct 97                           \cr
\end{tabular}
\end{table}

\end{document}